\documentclass[review]{elsarticle}

\usepackage{hyperref}

\journal{Journal of Magnetic Resonance}

\begin{document}

\begin{frontmatter}

\title{Extreme nuclear magnetic resonance: zero field, single spins, dark matter...}

\author{Dmitry Budker}
\address[mymainaddress]{Helmholtz Institute, Johannes Gutenberg-Universit\"{a}t Mainz, 55099 Mainz, Germany}
\address[mysecondaryaddress]{Department of Physics, University of California at Berkeley, Berkeley, CA 94720-7300, USA}

\date{\today}

\begin{abstract}
An unusual regime for liquid-state nuclear magnetic resonance (NMR) where the magnetic field strength is so low that the $J$-coupling (intramolecular spin-spin) interactions dominate the spin Hamiltonian opens a new paradigm with applications in spectroscopy, quantum control, and in fundamental-physics experiments, including searches for well-motivated dark-matter candidates. An interesting possibility is to bring this kind of ``extreme NMR'' together with another one---single nuclear spin detected with a single-spin quantum sensor. This would enable single-molecule $J$-spectroscopy. 
\end{abstract}


\end{frontmatter}


\section{Nuclear magnetic resonance without magnets?}
While the progress of NMR is often associated with the advent of better magnets with improved characteristics such as higher field, higher field homogeneity in a larger volume, or higher field gradients as in the case of magnetic resonance imaging (MRI), recent years have seen the emergence of zero- to ultralow-field (ZULF) NMR sometimes referred to as ``NMR without magnets''; see, for example, review \cite{Blanchard2016emagres}. Not only is there no need for magnets in ZULF NMR experiments, strong shielding from the Earth and laboratory magnetic fields is typically required \cite{Tayler2017}.

\begin{figure}[h!]
    \begin{center}
    \includegraphics[height=3. in]{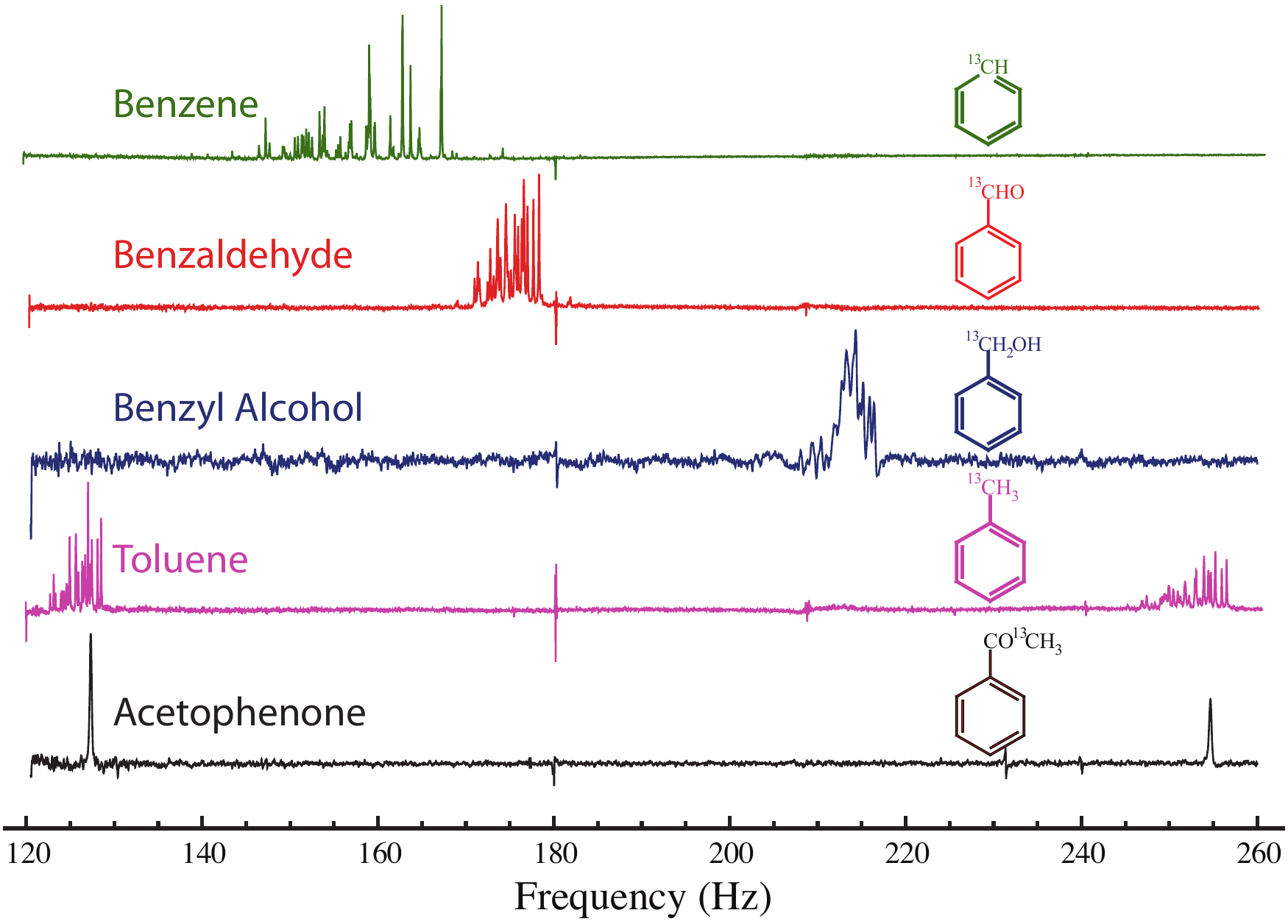}
\caption{Zero-field NMR spectra of various aromatic compounds \cite{Blanchard2013JACS} are completely different from one molecule to another. The complex splitting patterns of the spectral lines are due to coupling between distant spins and are largely reproduced or predicted with density-matrix modeling. The features at the harmonics of 60\,Hz are due to spurious power-line interference.}
    \label{fig1}
    \end{center}
\end{figure}

In order to explain how zero-field NMR is possible \cite{Ledbetter2013}, it is useful to begin by recalling exactly why magnetic fields are normally needed. An NMR experiment can be thought of as a sequence of spin polarization, encoding, and detection stages, with each of these stages normally depending on the presence of a magnetic field. The idea of zero-field NMR is to eliminate the need for magnets in every one of these three stages. We briefly discuss each of these.
\paragraph{Polarization} Thermal polarization, which is used in conventional NMR, is proportional to the magnetic field but is not the only or necessarily the best way to polarize nuclei. Methods like spin-exchange optical pumping (SEOP) or parahydrogen-induced polarization (PHIP) can produce orders of magnitude higher polarization and do not require strong (or any) magnetic fields. Another technique that does not rely on magnetic field is stochastic polarization (a.k.a. the ``do nothing technique''): an ensemble of $N$ randomly prepared spins will have an excess polarization on the order of $\pm \sqrt{N}$ spins along any direction. The method becomes essentially $100\%$ efficient as $N\rightarrow 1$. This is the polarization technique of choice in single-molecule NMR, see below.
\paragraph{Encoding} The role of magnetic fields and gradients is different for the two principal NMR modalities, spectroscopy and imaging. In the case of spectroscopy, chemical differentiation is typically based on the difference of chemical shifts for a given type of nuclear spins in different chemical environments. The role of the field is to ``pull apart'' the lines corresponding to different chemical shifts (measured in units of ppm) to better resolve and measure these. Typically smaller than chemical shifts and shift anisotropies, are couplings among nuclear spins, the $J$-couplings, that arise due to second-order hyperfine interactions. In contrast, in ZULF NMR, $J$-couplings are the dominant interactions, where in the ``ultralow-field'' regime \cite{Appelt2010}, Zeeman energies are smaller than $J$-couplings. A welcome feature of the ZULF regime from the point of view of spectroscopy is that the spectra are often completely different from one molecule to another (Fig.\,\ref{fig1}) as opposed to relatively small differences due to different chemical shifts. 
Only microtesla-level or even smaller fields are needed in the ZULF regime to produce Zeeman splitting of the $J$-coupled energy-level manifolds, which provides an additional handle on spectroscopy \cite{Ledbetter2011near}. The ZULF regime also allows one to measure non-secular spin couplings that are truncated in the presence of a strong magnetic field \cite{Blanchard2015}.

It should be mentioned that some of the benefits of zero-field encoding were exploited already in the original ``zero-field NMR'' (\cite{Zax85} and references therein), where polarization and detection were, in fact, performed in high field. 
\paragraph{Detection} While Faraday induction, where the signal with a given sample magnetization scales roughly as $B$, is used for detection in most NMR setups, there exist highly efficient magnetometers capable of detecting nuclear magnetization at arbitrarily low applied fields. Many recent ZULF NMR experiments have been done with atomic-vapor optical magnetometers  \cite{budker2013optical}; whereas some of the early ZULF NMR work employed superconducting quantum interferometers (SQUIDs) \cite{McDermott2002}.

Bringing all of this together, a complete ``NMR without magnets'' experiment becomes possible \cite{Theis2011}.

\section{Anticipated and unexpected applications}
While analytical applications were the initial motivation for the development of ZULF NMR, it was quickly realized that there is also a potential for fundamental-physics experiments \cite{Ledbetter2012liquid,Wu2018}, including searches for so-called ``ultralight bosonic dark matter'' that has already produced significant results \cite{Garcon2019,Wu2019} thus far corresponding to exclusion of certain parts of the plausible dark-matter parameter space. Another initially unanticipated application is robust quantum control and implementation of quantum gates in the ZULF regime \cite{Jiang2018}.

An example of a recent application of ZULF NMR in chemistry and biomedicine is the study of systems undergoing chemical exchange \cite{Barskiy2019}, including that of pyruvic acid, a molecule of importance for detecting cancer. 

\section{Color centers in diamond and single-spin NMR}
In the meantime, the past decade has seen rapid development of magnetic-sensing technology based on nitrogen-vacancy (NV) color centers in diamond \cite{Doherty2013}, offering an unprecedented combination of high sensitivity and spatial resolution \cite{Taylor2008}. If one uses a single NV center located a few nanometers below a diamond surface, such magnetometer can have nanometer-scale spatial resolution and enough sensitivity to detect not only single electron spins but even single nuclear spins located near the diamond surface \cite{Muller2014,Suter2017}, enabling a new exciting research area of single-spin NMR and MRI \cite{Sushkov2014}.

\section{Single-molecule ZULF NMR}
\begin{figure}[h!]
    \begin{center}
    \includegraphics[height=3. in]{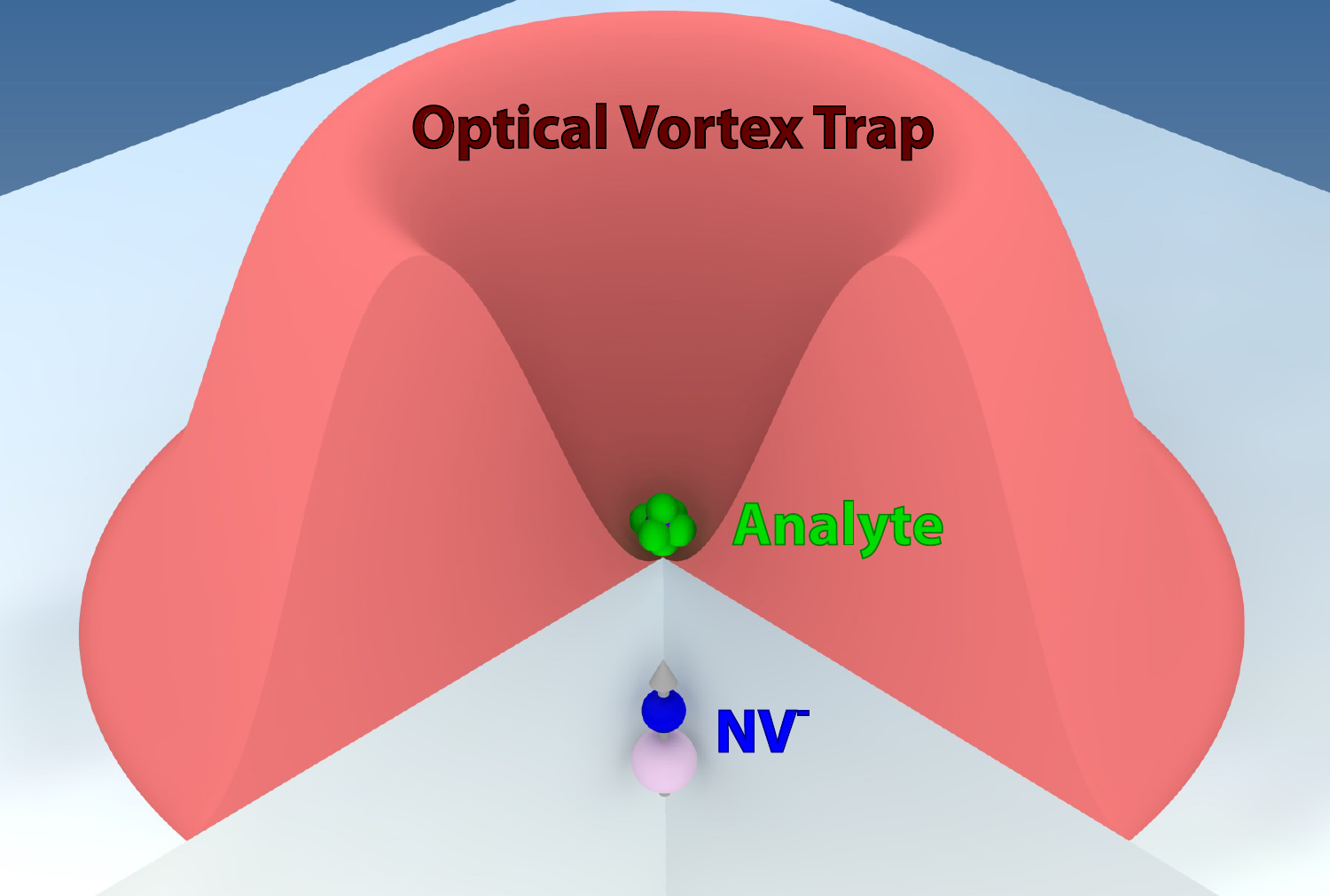}
\caption{Single-molecule NMR spectroscopy using a single NV center located a few nanometers below the surface of a diamond. The figure illustrates a possible technique for localizing the molecule close to the NV center using a ``doughnut'' shaped (vortex) optical beam similar to the ones employed in ``superresolution'' optical microscopy. Figure courtesy John W.\,Blanchard.}
    \label{fig2}
    \end{center}
\end{figure}

Combining the various ``NMR extremes'' discussed above, one arrives at the concept of single-molecule ZULF NMR with a stochastically polarized sample producing a signal detected with a single NV center \cite{McGuinness2019}, which is a goal of a current experimental effort. There are many challenges to be addressed in this program. For example, for $J$-spectroscopy not to be obscured by dipole-dipole interactions, the analyte molecule should be free to tumble. On the other hand, a free molecule would rapidly (on nanosecond time scales) diffuse across the several-nm region where the NV is sensitive  leading to excessive broadening of the resonance lines. Various approaches for confinement of ``free molecules'' within the sensitive region are being explored, with optimistic outlook, including molecular cages, ``tethering,'' optical confinement (Fig.\,\ref{fig2}), etc. 

\section{Predictions are hard...}
While such ``extreme NMR'' is undoubtedly elegant, we strongly believe it will also be useful. Not only will it allow investigation of samples with only single molecules available (for example, compounds containing superheavy elements) or measurements allowing to resolve molecular conformation dynamics at a single-molecule level, such measurements will also enable both the study and utilization of quantum effects that are already being explored with ZULF NMR at the ensemble level \cite{Jiang2018}.

In the realm of biology, one can imagine single-molecule ZULF NMR being useful for metabolomics at a single-cell level, allowing diagnostics of abnormalities at early stages of embryonic evolution or perhaps a massively parallel disease detection (and treatment) throughout a human body at such early stages that only a few malignant or abnormal cells are present. 

As with many disruptive technologies, the most exciting applications will likely be those that we cannot even imagine yet.

The author is indebted to D.~Barskiy, J.~Blanchard, P.~Bl{\"u}mler, F.~Jelezko, I.~Koptyug, T.~Lenz, L.~McGuindess, A.~Pines, and A.~Trabesinger for collaboration and comments on the manuscript. This work was supported by NSF (CHE-1709944), by the EU FETOPEN
Flagship Project ASTERIQS (action 820394) and
the German Federal Ministry of Education and Research
(BMBF) within the Quantumtechnologien program
(FKZ 13N14439 and FKZ 13N15064) and the Cluster
of Excellence Precision Physics, Fundamental Interactions,
and Structure of Matter (PRISMA+ EXC
2118/1) funded by the German Research Foundation
(DFG) within the German Excellence Strategy (Project
ID 39083149).

\bibliography{ZULF}

\end{document}